\documentclass{aa}  

\newcommand{\beq}[1]{\begin{equation}\label{#1}}
\newcommand{\eeq}{\end{equation}}
\newcommand{\sub}[1]{_{\rm #1}}

\newcommand{\Ms}{M\sub{s}}

\newcommand{\ap}{a\sub{p}}
\newcommand{\astop}{a\sub{\mathrm{stop}}}

\newcommand{\Mp}{M\sub{p}}

\newcommand{\Rp}{R\sub{p}}
\newcommand{\Pp}{P\sub{p}}
\newcommand{\Ps}{P\sub{s}}

\newcommand{\roche}{r\sub{R}}
\newcommand{\hill}{r\sub{H}}

\newcommand{\rev}[1]{\textcolor{black}{ #1}}
\newcommand{\revv}[1]{\textcolor{black}{ #1}}
\newcommand{\bea}{\begin{eqnarray}}
\newcommand{\eea}{\end{eqnarray}}

\usepackage{orcidlink}

\usepackage{natbib}

\usepackage{url}
\usepackage{soul}
\usepackage{multirow} 
\usepackage{makecell} 
\usepackage{booktabs}

\usepackage{txfonts}
\usepackage{graphicx}
\usepackage{txfonts,textcomp}

\usepackage{natbib,twoopt}
\bibpunct{(}{)}{;}{a}{}{,}
\usepackage[hyphenbreaks]{breakurl}
\usepackage{hyperref}

\hypersetup{
  colorlinks,
  citecolor=cyan,
  linkcolor=magenta,
  urlcolor=teal,
}

\begin{document} 

\title{Extreme exomoons in WASP-49 Ab: dynamics and detectability}
\author{Mario Sucerquia\inst{1}\thanks{E-mail: \href{mailto:mario.sucerquia@univ-grenoble-alpes.fr}{mario.sucerquia@univ-grenoble-alpes.fr}}\orcidlink{0000-0002-8065-4199}
        \and
        Nicolás Cuello\inst{1}\orcidlink{0000-0003-3713-8073}}
\institute{\centering
    Univ. Grenoble Alpes, CNRS, IPAG, 38000 Grenoble, France
}

   \date{\today}

 
\abstract
{WASP-49Ab, a low-density, Saturn-like planet in a tight orbit around a Sun-like star within a wide binary system, is a compelling candidate for hosting a volcanic moon, as suggested by the detection of Doppler-shifted sodium.
}
{This study evaluates the stability of potential satellites around WASP-49Ab under the influence of planetary oblateness, relativistic effects, and perturbations from a close companion star, focusing on their impact on light curve parameters such as transit duration and impact parameter variations, driven by the evolution of the planet's orbit in this extreme environment.
}
{Using N-body simulations and semi-analytical methods, we analysed \revv{moon's dynamics} across varied initial conditions and gravitational frameworks including the potential of an oblate planet and the effects of the general relativity.}
{We find that `selenity', a moon survival indicator, is high in close orbits with low eccentricity, near the Roche limit, especially for masses greater than Io's. Stability decreases as eccentricity or distance from the planet increases. Additionally, we observe a strong destabilising resonance near $1.4 \, \Rp$ when planetary eccentricities are considered to be $\gtrsim 0$.
}
{This study confirms the potential for stable exomoons around WASP-49Ab despite its hostile environment, emphasizing the importance of incorporating diverse physical effects in stability analyses, aiding future detection efforts.}


\titlerunning{WASP-49Ab, an extreme exomoon}

\authorrunning{Sucerquia \& Cuello. }
\maketitle
%

\section{Introduction}
\label{sec:intro}

According to formation models \citep{Canup2006}, moons are anticipated to be common around Jupiter-like planets. Given the prevalence of these planets, the successful detection—or absence—of exomoons would have significant implications for current models of planetary migration and lunar habitability \citep{Barnes2002, Sucerquia2019, Martinez2019, Sucerquia2020}.

Direct exomoon detection is challenging due to limited resolution and sensitivity to weak signals from small, nearby moons. A promising alternative is analysing light curve variations during planetary transits. These variations include Transit Timing Variations (TTVs, \citealt{Kipping2009a}), Transit Duration Variations (TDVs, \citealt{Kipping2009b}), and Transit Radius Variations (TRVs, \citealt{Rodenbeck2020}), which can provide insights into the presence of moons, or nearby planets. While difficult to detect, these secondary effects currently represent some of the most effective techniques for revealing the presence of undiscovered moons. 

The search for exomoons has gained significant interest, particularly with candidates like Kepler-1625b-i, a Neptune-sized moon orbiting a super-Jovian planet, and Kepler-1708b-i, approximately 2.6 Earth radii, orbiting its Jupiter-sized host at ~12 planetary radii. Kepler-1625b-i was proposed through mutual transits of the planet and moon \citep{Teachey-Kipping-2018}, while Kepler-1708b-i emerged from a survey of 70 cool giant exoplanets \citep{Kipping2022}. However, these findings are debated due to uncertainties in the physical and orbital characteristics of these systems \citep{Kreidberg2019}. The significant high planet-to-moon ratio of these candidates challenge traditional moon formation models, such as accretion capture, suggesting a new class of gas giant moons \citep{Heller2018, Teachey-Kipping-2018}, or even, ringed moons \citep{Sucerquia2022}.

An alternative approach to exomoon detection focuses on the activity within planetary environments. Stellar winds can erode nearby moons, producing halos of material, or tidal dynamics can expel particles, creating detectable photometric signatures around planets with moons. This is the case with the WASP-49Ab system \citep{Lendl2012}, a Saturn-like planet in a 2.8-day S-type orbit around a Sun-like star within a binary system, where neutral sodium (Na I) has been detected in the planetary atmosphere \citep{Oza2024}. The detection of this alkali metal, with notable night-to-night variations in spectral flux, suggests the possible existence of an external source, such as a volcanic exomoon, akin to Jupiter’s Io. Observations with the VLT/ESPRESSO reveal a Doppler shift of approximately +9.7 km/s relative to the planet’s reference frame, which could suggest the presence of a natural satellite orbiting WASP-49Ab.

\rev{Potential moons around WASP-49Ab must reside in a narrow region between the Roche limit and the Hill sphere ($\sim1.2$--$1.8$ planetary radii) to maintain dynamical stability, avoiding destabilising evection and eviction resonances \citep{Vaillant2022}. Whilst previous studies have explored individual perturbations, such as tides \citep{Barnes04, Domingos2006, Alvarado2017} or binary companions \citep{Quarles2021, Gordon2024}, here we integrate multiple effects—including planetary oblateness \citep{Hong2015} and relativistic corrections, making the inclusion of these under-explored effects pivotal in this extreme environment.
}

\revv{We first describe the WASP-49Ab system in Sec.~\ref{sec:WASP-49Ab}. Then, we report our numerical simulations to assess the long-term stability of potential satellites and examine the impact on light curves in Secs. ~\ref{sec:metodology} and ~\ref{sec:results}. Finally, we discuss how these perturbations may impact upcoming exomoon detections in Sec.~\ref{sec:discussion}.}

\vspace{-0.2cm}

\section{The WASP-49Ab system and perturbative forces}\label{sec:WASP-49Ab}

WASP-49Ab is a hot Saturn-like exoplanet with a mass of 0.37 times Jupiter's mass ($M_J$) and a radius ($\Rp$) of 1.19 times Jupiter's radius ($R_J$), orbiting the G6V star WASP-49A at 0.03 AU \citep{Lendl2012}. This proximity to its host star results in significant stellar irradiation, and possibly places the planet in a tidally locked configuration due to its short orbital period of 2.8 days. WASP-49Ab is also part of a wide binary system, with a distant K-dwarf companion located 443 AU away. 

High-resolution transmission spectroscopy has detected variable sodium absorption at significant altitudes in WASP-49Ab's atmosphere \citep{Wyttenbach2017}. Observations indicate potential sodium flux variability, possibly driven by external sources, such as volcanic activity from a nearby exomoon, analogous to the volcanic interactions observed in the Jupiter-Io system. \cite{Oza2024} provided further evidence, detecting a redshifted sodium signature at +9.7 km/s, consistent with material potentially originating from a satellite orbiting between the planet’s Roche limit and Hill radius.

\vspace{-0.2cm}
\subsection{Orbital constraints for potential satellites}\label{subsec:OrbitalConstraints}

\rev{Stable orbits for potential satellites around WASP-49Ab are expected to exist within a narrow region bounded by the Roche limit (\(\roche\)) and the Hill radius (\(\hill\)). The Roche limit represents the minimum distance at which a satellite can avoid tidal disruption by the planet. For a satellite with mass \(M_\mathrm{s}\) and radius \(R_\mathrm{s}\), \(\roche\) is given by:}
\begin{equation}
    \roche = 2.44 \left( \frac{\Mp}{M_\mathrm{s}} \right)^{\frac{1}{3}} R_\mathrm{s},
    \label{eq:roche}
\end{equation}
where $\Mp$ is the planet’s mass. Within this limit, the tidal forces would exceed the moon’s self-gravity, leading to disintegration.

The outer stability boundary is defined by the secondary Hill radius, \(4.8 \, r_\mathrm{H}\) for prograde orbits \citep{Domingos2006}. The Hill radius (\(r_\mathrm{H}\)) defines the region where the planet's gravitational influence dominates over that of the star, calculated as:
\begin{equation}
    r_\mathrm{H} = \ap \left( \frac{\Mp}{3 M_\star} \right)^{\frac{1}{3}},
    \label{eq:hill_radius}
\end{equation}
where \(\ap\) is the semi-major axis of the planet’s orbit around the star and \(M_\star\) is the mass of the host star. For WASP-49Ab, this region extends up to approximately \(3.4 \, \Rp\), beyond which the star's gravity would likely eject the satellite, by mean of evection and eviction resonances that affect the satellite eccentricity and inclination, respectively \citep{Vaillant2022}.

Thus, the stable orbital region for potential satellites around WASP-49Ab is constrained between the Roche limit and the secondary Hill radius, forming a delicate balance where satellites could survive the strong tidal and gravitational forces from both the planet and the host star.
\vspace{-0.2cm}

\subsection{Perturbative forces and orbital evolution of satellites} \label{subsec:PerturbativeForces}

\rev{Satellite stability in this narrow region and extreme environment is influenced by various perturbative forces, which affect orbital parameters over different timescales. In the following, we detail these effects (see also Appendix~\ref{ap:timescales}).}
\vspace{-0.2cm}

\paragraph{Stellar and planetary tides:} Tidal interactions from both the host star and the planet play a crucial role in shaping the orbital evolution of satellites \rev{ and planets, particularly in close-in systems like WASP-49Ab \citep[see, e.g.,][and referencer therein]{Makarov2023}}. Stellar tides induce bulges on the planet, leading to energy dissipation and a gradual evolution of the satellite's orbit. \rev{The rates of change in the semi-major axis and orbital period due to these tidal forces are described in equations~ 9 and 10 in \citet{Alvarado2017} --- based on the framework by \citet{Barnes2002} and the classical tidal dissipation model of \citet{Goldreich1966}, for a system where the planet’s physical characteristics evolve over time.}
 Here, we assume the moon has survived any detachment processes and has reached its asymptotic semi-major axis, \(a_{\text{stop}}\), defined by  \citet{Sucerquia2020} as the orbital distance beyond which tidal forces no longer affect the moon’s orbit. This model assumes a slowly evolving planetary interior, but the planet’s close proximity to its star could result in elevated temperatures, affecting its size, internal structure, which should affect the long-term stability of the satellite’s orbit.
\vspace{-0.5cm}
\rev{\paragraph{Perturbations from an outer body:} In non-coplanar three-body systems, secular perturbations, including von Zeipel-Kozai-Lidov (ZLK) oscillations and periapsis precession \citep{vonZeipel1910,Lidov1962,Kozai1962,Hamers2018b}, can significantly impact long-term stability. In the WASP-49 system, these perturbations act through (1) interactions between the planet and WASP-49B, affecting the planet’s orbital parameters over tens of thousands of years, and (2) interactions between the satellite, the planet, and the primary star, influencing the satellite’s orbit on shorter timescales (weeks to months). For WASP-49Ab, ZLK effects driven by the binary companion are negligible due to the large separation and weak planetary perturbations. Similarly, periapsis precession of the planet due to the stellar companion should occur at a slow rate (\(\sim10^{-7} \, \mathrm{rad/yr}\)) and has negligible impact on satellite stability. However, stellar-induced ZLK effects on satellites with inclinations (\(i > 39^\circ\)) could cause destabilising eccentricity and inclination oscillations, making them highly prone to instability within 1.2–1.8 \(\Rp\). Consequently, such high-inclination satellites are excluded from this study.}
\vspace{-0.5cm}
\paragraph{Planetary oblateness:} Due to its close proximity to the host star, WASP-49Ab is presumed to be tidally locked, meaning that its rotational period is synchronised with its orbital period. This tidal locking promotes a stable orientation, leading to an enhanced equatorial bulge, or oblateness, which introduces additional gravitational perturbations affecting the orbits of any satellites. The gravitational influence of planetary oblateness is quantified by the gravitational harmonic coefficients \(J\). The dominant term, \(J_\mathrm{2}\) (quadrupole moment), represents the primary deviation from a perfect sphere and significantly affects satellite dynamics by inducing precession of the satellite’s argument of periastron, \(\omega\), with an effect that weakens with distance. In systems with an oblate host planet, the increased nodal precession rate induced by $J_\mathrm{2}$ stabilises the satellite’s orbit by preventing secular resonances with the host, which could otherwise destabilise the satellite’s inclination and eccentricity. This effect, as demonstrated by \cite{Hong2015}, mitigates the destabilising influence of the moon’s inclination and enhances overall orbital stability relative to configurations with a spherical host planet. In this analysis, we include only the \(J_\mathrm{2}\) term to account for the primary stabilising effect of planetary oblateness on the satellite’s orbit. The rate of change in the planet's periastron \(\omega_{\mathrm{J}}\), due to \(J_\mathrm{2}\) is expressed as:
\begin{equation}
    \frac{\mathrm{d}\omega_{\mathrm{J}}}{\mathrm{d}t} = \frac{3}{2} J_\mathrm{2} \left(\frac{R_p}{\ap}\right)^2 \frac{\sqrt{G\Mp}}{\ap^{7/2}},
    \label{eq:precession}
\end{equation}
where \( R_\mathrm{p} \) denotes the planet's radius, \( a_\mathrm{p} \) its semi-major axis, \( M_\mathrm{p} \) its mass, and \( G \) is the gravitational constant. This precession effect can lead to significant changes in the orbital orientation of close-in satellites over time, contributing to the long-term stabilisation of their orbits by damping inclination and eccentricity variations. 
To calculate the shape coefficient $J_\mathrm{2}$ it was assumed that its rotation period equals its orbital period (\( P_{\text{rot}} = P_{\text{orb}} \)). With this assumption, the angular velocity \(\Omega\) was calculated as \( \Omega = \frac{2 \pi}{P_{\text{rot}}} \). Then, we obtain \( J_\mathrm{2} \) using the following equation: 
\begin{equation}
J_\mathrm{2} = \frac{3}{4} \frac{\Omega^2 \, \Rp^3}{G \, \Mp} 
\end{equation}
where $\Mp$ and $\Rp$ are the planet's mass and radius, respectively. With an estimated \( J_\mathrm{2} \approx 0.0063 \) ---indicating greater oblateness than Earth but less than Jupiter--- this coefficient suggests a substantial equatorial bulge driven by centrifugal forces, resulting in a pronounced deviation from sphericity. Such oblateness could notably impact the planet’s gravitational field, its dynamical evolution, and the orbital dynamics of nearby satellites.
\vspace{-0.2cm}
\paragraph{Relativistic effects:} Due to the close proximity of WASP-49Ab to its star, relativistic effects introduce an additional rate of precession, $\omega_{\mathrm{GR}}$, described by:
\begin{equation}
    \frac{\mathrm{d}\omega_{\mathrm{GR}}}{\mathrm{d}t} = \frac{3  G  M_\star n_\mathrm{p}}{\ap \; (1 - e_\mathrm{p}^2) \; c^2},
    \label{eq:relativistic}
\end{equation}
where \( M_\star \) is the stellar mass, \( e_\mathrm{p} \) denotes the planet's orbital eccentricity, \( c \) is the speed of light, and \( n_\mathrm{p} = 2\pi/P_\mathrm{p} \) is the planet’s mean orbital frequency, with \( P_\mathrm{p} \) representing the orbital period. This relativistic correction leads to a continuous, gradual shift in the orientation of the planet’s elliptical orbit around its star. Although the effect is subtle per orbit, it accumulates significantly over many orbital cycles, resulting in a measurable advance of the pericentre position \citep[See e.g.][]{Jordan2008}. 


\vspace{-0.2cm}
\section{Methodology}\label{sec:metodology}

\subsection{Dynamical evolution}

We investigated the stability of hypothetical satellites around WASP-49Ab using the {\tt REBOUND}  N-body code \citep{rebound}, with additional physics from the {\tt REBOUNDx} package \citep{reboundx}. The system comprises a star, a planet, and an array of non-mutually interacting moons, configured to explore a wide range of orbital and physical parameter spaces. These satellites experience perturbations as outlined in Sec. \ref{sec:WASP-49Ab}. Specifically, we calculated the \revv{Mean Exponential Growth Factor of Nearby Orbits} (MEGNO) parameter to get an insight on the general dynamical features system \citep{Cincotta2000}.

\revv{Additionally, we assessed the likelihood of satellite retention under various dynamical conditions, incorporating first-order relativistic corrections and planetary oblateness up to \( J_\mathrm{2} \). This analysis led to the development of a metric, termed `selenity', which quantifies a satellite's ability to remain bound to its planet despite external perturbations. The metric is derived from simulations across the \((a_\mathrm{s}, e_\mathrm{s})\) or \((a_\mathrm{s}, m_\mathrm{s})\) parameter spaces, evaluated over varying inclinations. For each point, the ratio of stable outcomes (bound satellites) to total cases is computed, yielding a normalised value between 0 and 1. This ensures that the satellite’s semi-major axis remains within the primary Hill radius, its eccentricity stays between 0 and 1, and it avoids collisions with the planet. Collectively, these criteria establish whether the satellite maintains a stable, bounded orbit within the system's dynamical constraints.}

To compute the Roche's radius (Eq. \ref{eq:roche}) for the satellite radius, we adopted a mass-radius scaling relation applicable to rocky moons, with the form \( R \propto M^{1/3.5} \), based on Io's characteristics but without adjustments for core mass fraction variations \citep{Sotin2007}. This approach provides a representative scale for rocky bodies similar to Io, without specific composition dependence. We set the eccentricity of WASP-49Ab to an average value of 0.03, given the range variation between 0.0 and 0.05, based on the posterior probability distribution obtained from the MCMC analysis done by \citealt{Wyttenbach2017}. \rev{Given the short orbital period of the planet and the anticipated orbital periods of potential moons on the scale of hours, we selected the Hamiltonian {\tt WHFast} integrator for the system's evolution, as close encounters are not a primary focus. A timestep of \(10^{-5} \, \mathrm{yr}\) was used, mapping the planet’s orbit with approximately 800 points and the moon's orbit with about 70 points per cycle. Each simulation was conducted over 1000 planetary periods, encompassing approximately $10^5$ orbits of a hypothetical moon with an orbital period of roughly 10 hours. This setup allows for a comprehensive exploration of satellite dynamics across a wide range of configurations. See Appendix~\ref{ap:timescales} for details on the timescales.}


\vspace{-0.3cm}
\subsection{Detectability of exomoons around WASP-49Ab}

\rev{A planet with a massive satellite orbits the centre of mass of the planet-satellite system, causing Transit Timing Variations (TTV) and Transit Duration Variations (TDV) due to changes in the planet’s position and velocity relative to this centre of mass during transits.} TTV signals are most pronounced for circular, coplanar orbits, and for edge-on systems. The root-mean-square (rms) TTV amplitude, \( \delta_\mathrm{TTV} \), and TDV amplitude, \( \delta_\mathrm{TDV} \), are given by \citep{Kipping2009b}: \begin{eqnarray}
\label{eq:dttv}
\delta_\mathrm{TTV} & = & \frac{1}{2 \pi} \frac{ \astop \,\Ms}{\ap \, \Mp} \Pp \sqrt{\frac{\Phi_\mathrm{TTV}}{2 \pi}}, \\
\label{eq:dtdv}
\delta_\mathrm{TDV} & = & \bar{\tau}  \frac{ \astop \,\Ms \Pp}{\ap \, \Mp \; \Ps} \sqrt{\frac{\Phi_\mathrm{TDV}}{2 \pi}}.
\end{eqnarray}
where \( \Phi_\mathrm{TTV} \to \pi \) for aligned, circular orbits. Moreover, $\bar{\tau}$ is the averaged transit duration given by
\begin{equation}
    \bar{\tau} = \frac{\Pp}{\pi} \arcsin{\left[\sqrt{\frac{(R_* + \Rp)^2 - (b R_*)^2}{\ap^2 - b^2 R_*^2 }}\right]},
    \label{eq:transdur}
\end{equation}
where \( \Ps \) is the moon’s orbital period, \( R_* \) is the stellar radius. Eqs.~\ref{eq:dttv} and \ref{eq:dtdv} do not depend on the moon’s apparent size, making them independent of secondary effects such as an atmosphere or extended material around the moon. However, a massive moon can produce significant TTV and TDV signals, with amplitudes directly proportional to the moon’s mass. Additionally, simultaneous transits of the moon can alter the perceived depth of the transit, either within a single transit or across multiple transits, depending on the orbital periods of the moon and the planet.

\vspace{-0.2cm}
\section{Results and analysis}\label{sec:results}

\subsection{Exomoon dynamics}
\label{sec:dyn}

\begin{figure*}
    \centering
    \includegraphics[width=0.44\textwidth]{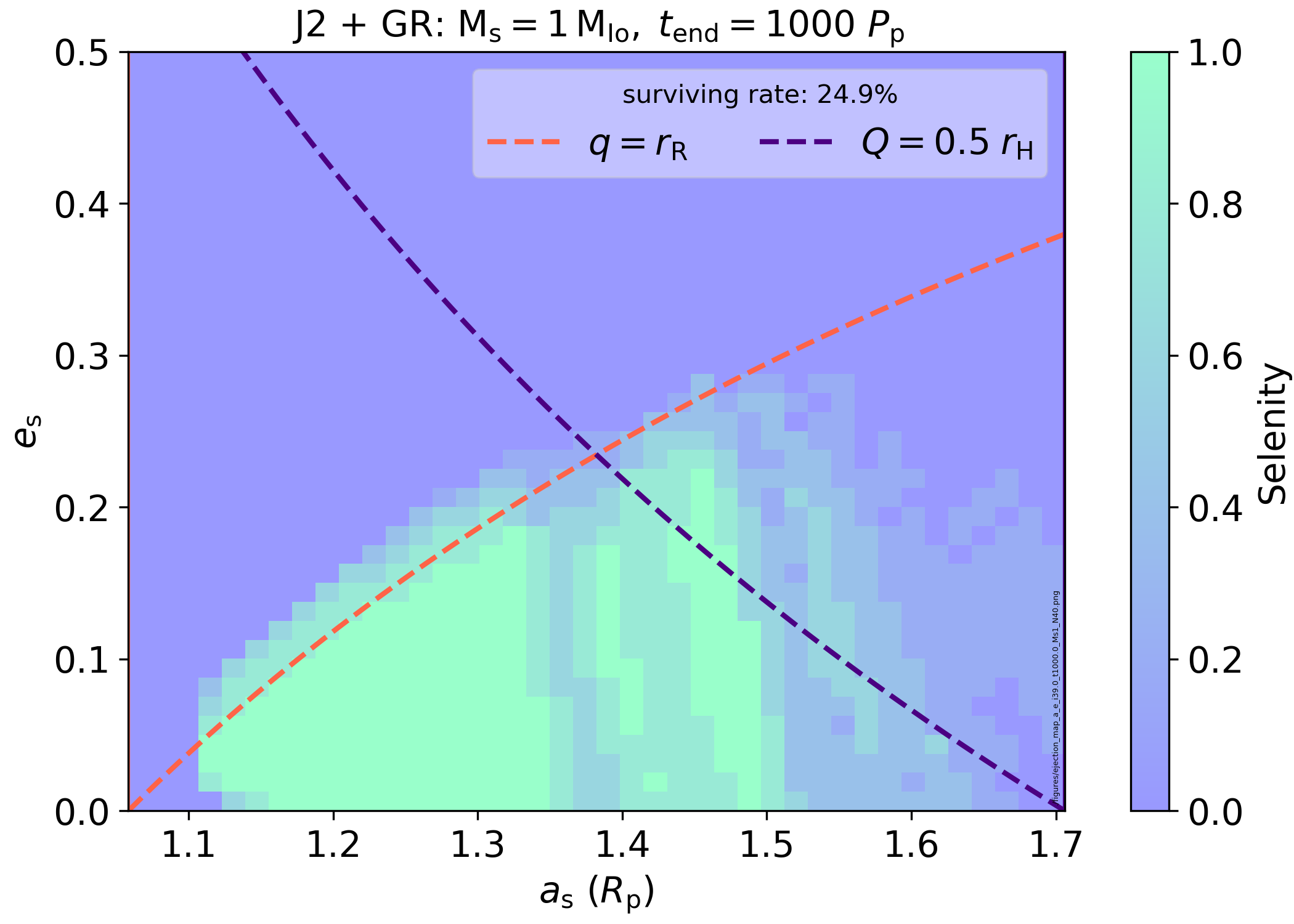}
    \includegraphics[width=0.44\textwidth]{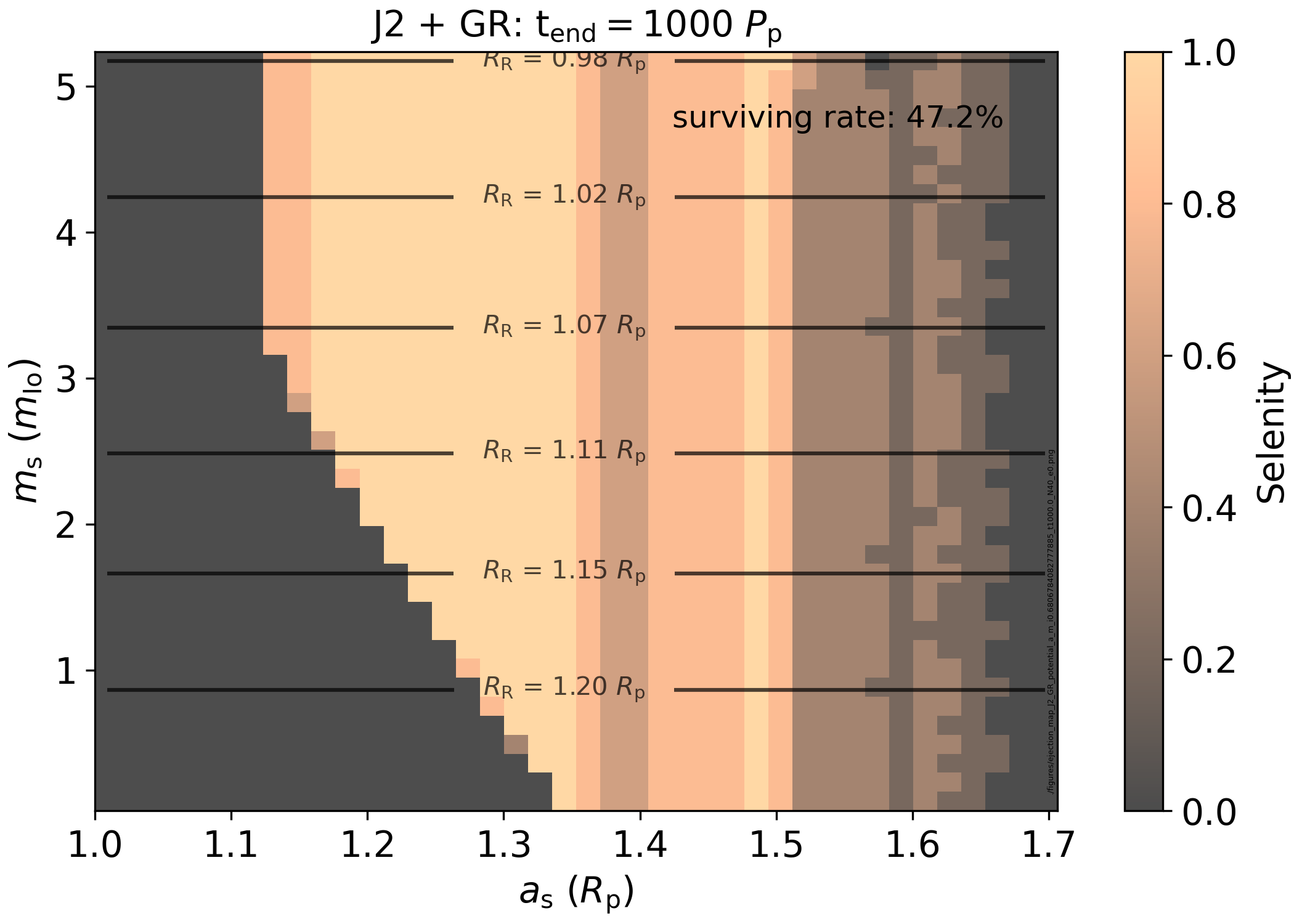}
    \caption{
    Stability maps illustrating survival and ejection probabilities for exomoons orbiting a close-in giant planet, factoring in the planet's \(J_\mathrm{2}\) quadrupole moment and relativistic corrections \rev{over 1000 $\Pp$}. The colour scale indicates `selenity’, showing the moon's likelihood of remaining bound to the planet under external perturbations. The left panel shows stability in the \(a_\mathrm{s} - e_\mathrm{s}\) plane, with the Roche limit and Hill radius as key markers. The right panel displays stability in the \(a_\mathrm{s} - M_\mathrm{s}\) plane, reflecting stability variations based on satellite mass and planet distance.
    }
    \label{fig:stability_maps}
\end{figure*}

In Fig.~\ref{fig:stability_maps}, we highlight regions where moons are most likely to remain stable. The colour scale in each panel represents the `selenity' metric, with a selenity value of 1 indicating strong gravitational binding to the planet and a value of 0 indicating that the satellite is unbound. Intermediate values suggest partial instability, primarily influenced by the satellite's initial inclination.

The left panel of Fig.~\ref{fig:stability_maps} shows the stability map in the \(a_\mathrm{s} - e_\mathrm{s}\) plane, where \(a_\mathrm{s}\) is the semi-major axis in units of planetary radii (\(\Rp\)) and the orbital eccentricity \(e_\mathrm{s}\). \rev{Each point in the simulation, representing a specific combination of semi-major axis and eccentricity, was evaluated across five different inclinations ranging from \(0^\circ\) to \(39^\circ\), ensuring that the effects of potentially destabilising resonances, such as ZLK resonances, were adequately accounted for and mitigated. The results were then combined ('stacked') and normalised to create a unified stability map.} Two key dynamical boundaries are highlighted: the red dashed line marking the Roche limit and the purple dashed line denoting the Hill radius. Light green regions indicate high orbital stability, where satellites are likely to remain bound, while blue regions represent areas with a higher probability of ejection. In this configuration, the calculated survival rate for satellites is 24.9\%. The right panel of Fig.~\ref{fig:stability_maps} displays the stability map in the \(a_\mathrm{s} - m_\mathrm{s}\) plane for circular orbits, where \(m_\mathrm{s}\) is the satellite’s mass in units of Io’s mass (\(m_{\text{Io}}\)). The light orange regions denote configurations with greater orbital stability, where satellites show a survival rate of 47.2\%, while the dark brown to black areas highlight regions of increased ejection probability. The labelled lines indicate various Roche radii, which vary with the satellite's mass. This could be seen as the minimum distances required to avoid tidal disruption. 
\vspace{-4mm}

These stability maps, which align well with and complement the MEGNO stability maps in Fig.~\ref{fig:MEGNO}, offer valuable insights into the survival constraints for satellites in the WASP-49Ab system, highlighting orbital configurations with increased resilience to ejection mechanisms. Note that the MEGNO indicator is designed for Newtonian dynamics, whereas `selenity' incorporates more complex interactions and environmental factors.
In summary, potential satellites in this system must reside very close to their planet, with low eccentricities (up to 0.2) and low inclinations. \rev{Furthermore, stability improves for satellites with masses similar to or greater than Io’s}. Masses below this threshold render the system unstable, while higher masses contribute to stability but cannot exceed five times Io’s mass to \rev{satisfy} the $M_\mathrm{s}/\Mp \leq 10^{-4}$ ratio found by \citealt{Canup2006}. 
\vspace{-0.4cm}
\subsection{Transit features and relativistic effects}
\begin{figure*}
    \centering
    \includegraphics[width=0.44\textwidth]{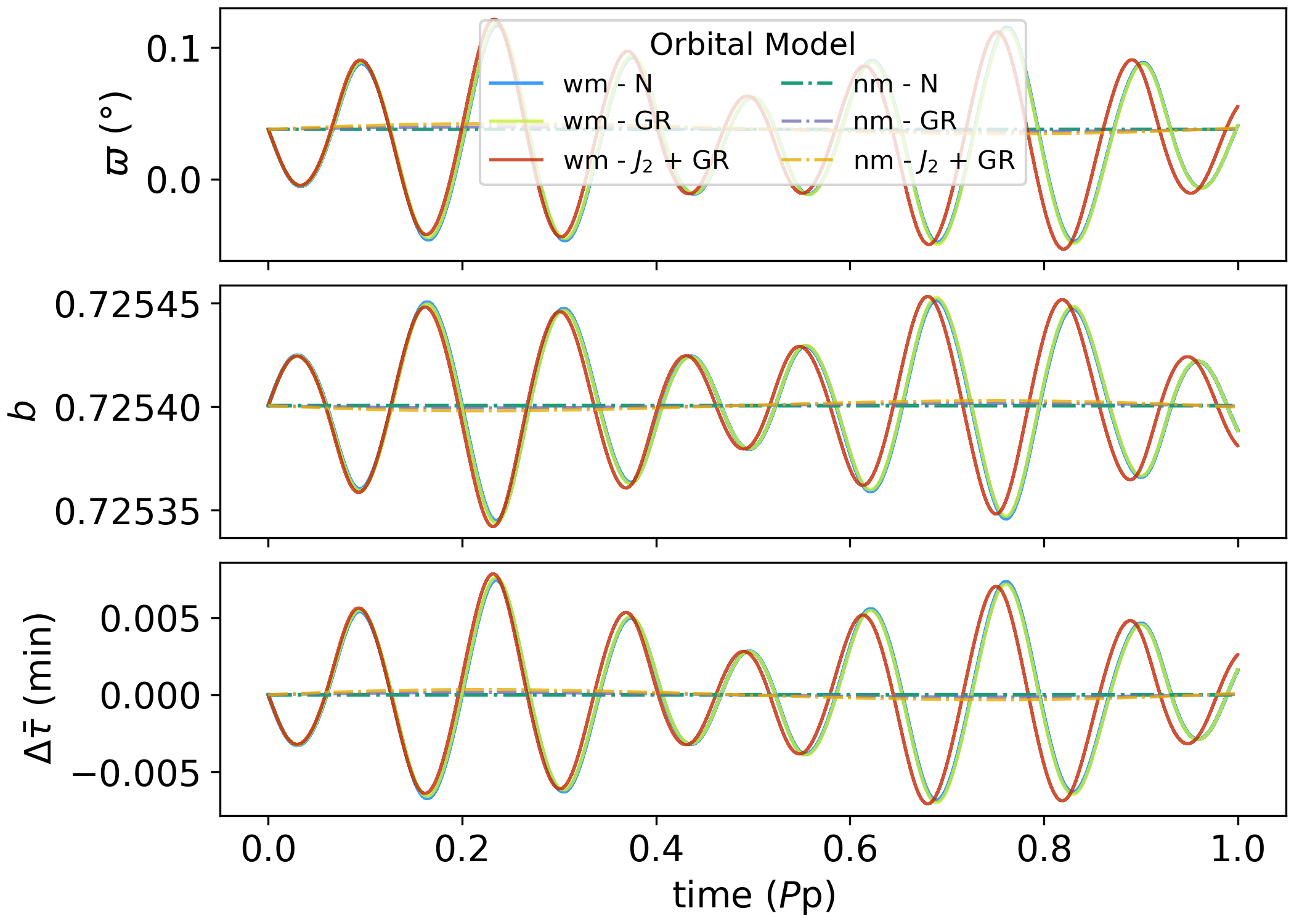}
    \includegraphics[width=0.44\textwidth]{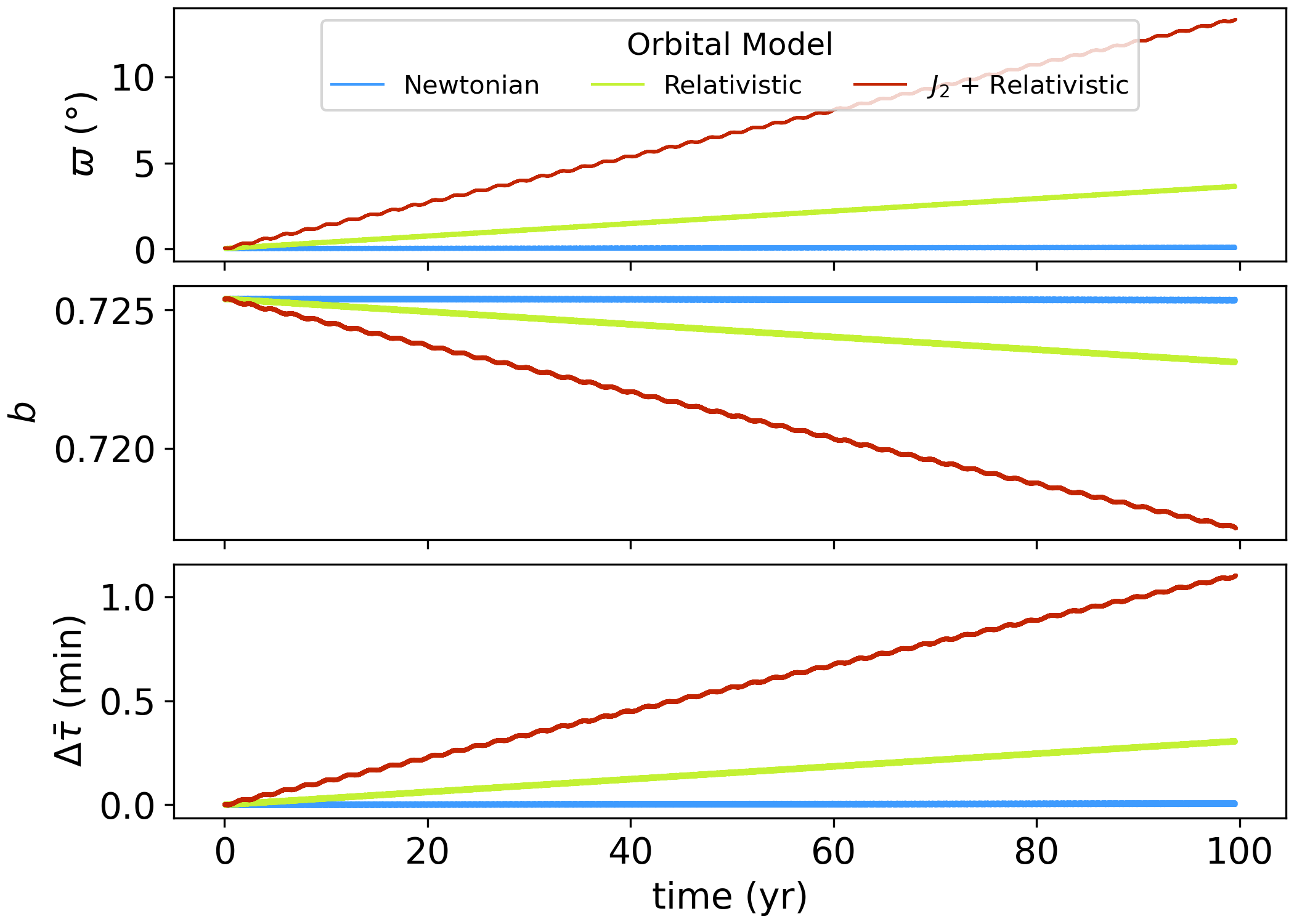}
    \caption{
     \rev{Evolution of the argument of pericentre ($\varpi_\mathrm{p}$), the impact parameter ($b$), and the transit duration ($\Delta \bar{\tau} = \bar{\tau}_f - \bar{\tau}_0$) for WASP-49Ab over a single planetary period (left panel) and over 100 years (right panel)}. The left panel illustrates the parameter evolution both with a moon (`wm', solid lines) and without a moon (`nm', dashed lines) under different orbital models: Newtonian (N), relativistic (GR), and relativistic with \(J_\mathrm{2}\) effects ($J_\mathrm{2}$ + GR). The coloured lines represent these models, highlighting the significant differences in the long and short-term dynamics of the system. 
    }
    \label{fig:transit_evolution}
\end{figure*}

The restricted range of values for the planet's mass and semi-major axis imposes significant constraints on its detectability via indirect dynamical methods. Applying Eqs.~\ref{eq:dttv} and \ref{eq:dtdv} to the most optimistic estimates for $m_\mathrm{s}$ and $a_\mathrm{s}$ from Sec.~\ref{sec:dyn}, the predicted transit timing and duration variations remain under a fraction of a minute. However, due to the extreme proximity of WASP-49Ab to its host star, relativistic precession and planetary deformation introduce a cumulative shift in the orientation of the orbital plane, which can affect the observed impact parameter and transit duration \citep{Jordan2008}, potentially mimicking satellite-induced variations and complicating interpretation.

The rate of periastron precession, influenced by both general relativistic effects and the planet's complex gravitational potential, affects \(\omega_\mathrm{GR}\) and \(\omega_\mathrm{p}\) over time, as given by Eqs.~\ref{eq:relativistic} and \ref{eq:precession}. This precession causes a gradual rotation of the orbital plane, impacting the impact parameter \( b \), defined as the projected distance between the centres of the planet and the star at mid-transit. Notably, \( b \) is particularly sensitive to this precession:
\begin{equation}
b = \frac{\ap \cos(i_\mathrm{p})}{R_{\star}} \frac{1 - e_\mathrm{p}^2}{1 + e_\mathrm{p} \sin \omega_\mathrm{p}},
\end{equation}
where \( R_{\star} \) denotes the stellar radius, and \( i \) is the orbital inclination. As the argument of periastron \(\omega_\mathrm{p}\) progresses, slight variations in \( b \) occur, modulating the transit geometry and consequently affecting the transit duration. Last, the transit duration $\bar{\tau}$, which depends on both the impact parameter and the orbital velocity of the planet at mid-transit, can be estimated by Eq.~\ref{eq:transdur}.

Over a timescale of 100 years ($\sim$13,000 orbits), we calculate a cumulative transit duration change, $\delta \tau$, of approximately 1.3 minutes for an Io analogue at 1.3 $\Rp$, resulting from relativistic and oblateness-induced precession. Fig. \ref{fig:transit_evolution} illustrates, in the left panel, the short-term evolution of $\varpi$, $b$, and $\bar{\tau}$ over time, while the right panel highlights the gradual drift in transit duration over a century. On shorter timescales (left panel), small shifts in frequency and amplitude become noticeable after a single planetary period across different models: Newtonian (solid lines) and relativistic (dashed lines), each evaluated with (`wm') and without (`nm') a massive moon. In the `nm' model, amplitudes are approximately an order of magnitude lower than in the `wm' model.
\revv{Although these rms amplitude variations are small (\(\delta b \sim 10^{-3}\) and \(\delta \bar{\tau} \sim 10^{-3} \, \text{min.}\)), they may still influence noise levels in detection efforts. Over longer timescales, however, the most significant effect appears in the full model, showing a cumulative shift in the argument of pericentre of approximately 13° per century.}

\vspace{-0.3cm}
\section{Discussion and conclusions} \label{sec:discussion}
\vspace{-0.1cm}
We investigated the evolution of hypothetical satellites around the close-in warm Saturn WASP-49Ab, considering its complex dynamical environment, including planetary deformation and general relativistic effects, within a numerical framework. Our focus was on satellites inside the narrow region between the Roche limit and the secondary Hill radius, constrained to moderate eccentricities and inclinations. Through a series of simulations assessing the MEGNO indicator and the `selenity’ metric (see Sec~\ref{sec:metodology}), we found that stability is favoured for moons with masses similar to Io. 

The MEGNO and `selenity' maps revealed a chaotic region around $a_s \approx 1.4\, \Rp$, initially suspected to result from mean-motion resonances. However, a detailed analysis of semi-major axes from $1.3$ to $1.5 \, \Rp$ found no significant mean-motion resonances up to the fifth order, suggesting that the chaos is not driven by such resonances. The `selenity’ metric further constrained the semi-major axis and eccentricity ranges for which moons remain bound to the planet, keeping the eccentricities of surviving moons below 0.22. Notably, even with significant inclinations, moons displayed considerable stability. This is attributed to the oblate shape of the host planet, which enhances the nodal precession rate, thereby mitigating destabilising secular resonances as was described by \cite{Hong2015}. Moreover, in simulations where \(e_\mathrm{p}\) was initialised to zero, the chaotic behaviour was eliminated, indicating that secular precession effects induced by the planet’s eccentricity are the primary cause, likely arising from a \(\nu\)-type secular resonance.



From a detectability perspective, we found that the $\delta_\mathrm{TTV}$ and $\delta_\mathrm{TDV}$ signals are below tenths of a minute, making them undetectable with current instrumentation. However, relativistic effects and the planet’s rotational deformation significantly alter the rate of change in the planet's argument of pericentre, $\varpi_\mathrm{p}$. Over long timescales, this causes a gradual shift in the orbital orientation and modifies the transit duration. This effect is more pronounced for eccentric orbits, but for WASP-49Ab, we estimate a cumulative impact on transit duration of approximately 1.3 minutes per century. On shorter timescales, by contrast, we observed an oscillatory pattern in the planet's impact parameter, affecting its third decimal place. \rev{Such variations may affect parameter estimates across studies over different epochs.}

In summary, we find that --- even in extreme environments like that of WASP-49Ab --- exomoons can remain dynamically stable. \rev{We also demonstrated that} accurate exomoon detection requires accounting for various sources of gravitational perturbations, including stellar multiplicity, higher-order gravitational effects, and interactions with other planets or sibling satellites. This work provides a general framework to \rev{motivate, guide, and support the upcoming search of exomoons.} 



\begin{acknowledgements}
   This project was supported by the European Research Council (ERC) under the European Union Horizon Europe research and innovation program (grant agreement No. 101042275, project Stellar-MADE). We also thank MSP and the Stellar-MADE team for their kind support.
\end{acknowledgements}
\vspace{-0.4cm}
\bibliographystyle{aa}
\vspace{-0.4cm}
\bibliography{biblio}

\appendix

\section{Dynamical timescales}
\label{ap:timescales}
\begin{table*}
\centering
\caption{Dynamical effects and their associated timescales in the WASP-49Ab system.}

\footnotesize
\begin{tabular}{@{}lll@{}}
\toprule
\textbf{Dynamical Effect}          & \textbf{Description}                                                     & \textbf{Typical Timescale}               \\
\midrule
Orbital period of satellite        & Time taken by the satellite to complete one orbit around the planet.     & $[6.7$ – $11.3] \, \mathrm{h}$ 
 \\
Orbital period of planet           & Time taken by the planet to complete one orbit around the host star.     & $2.8 \, \mathrm{d}$ 
\\
Precession due to $J_\mathrm{2}$            & Precession of the satellite's orbital plane caused by the planet's oblateness. & $\sim 55 \, \mathrm{d}$ \\
ZLK (planet-satellite-star)             & Oscillations in inclination and eccentricity induced by the host star.   & $[17 \, - 30] \, \mathrm{d}$
 \\
Orbital period of binary           & Time taken by the stellar binary to complete one orbit.                  & $7.18 \times 10^3 \, \mathrm{yr}$ \\
Relativistic precession            & Precession of the planet’s periapsis due to general relativistic effects. & $\sim  10^4 \, \mathrm{yr}$ \\
ZLK (planet-binary)                & Oscillations in the planet’s inclination and eccentricity induced by WASP-49B. & $\sim  10^9 \, \mathrm{yr}$ \\
\bottomrule
\end{tabular}
\label{tab:dynamical_effects}
\tablefoot{Timescales are calculated for satellites located between the Roche limit and 1.5 planetary radii.}
\end{table*} 

\rev{The dynamical timescales of a hypothetical satellite in the WASP-49Ab system are governed by gravitational interactions, relativistic effects, and the planet's oblateness. These processes collectively determine the evolution of the satellite's orbit. Below, we summarise the origin of the timescales in Table~\ref{tab:dynamical_effects}.}

\revv{The timestep in our simulations is determined by the shortest dynamical timescale in the system. As shown in Table~\ref{tab:dynamical_effects}, the orbital period of the moon is the fastest process, ranging from approximately $7 \times 10^{-4}$ to $1.3 \times 10^{-3} \, \mathrm{yr}$ ($6$–$11 \, \mathrm{h}$). To ensure numerical stability and accuracy, we adopt a timestep of $10^{-5} \, \mathrm{yr}$ for the maps presented in Figs.~\ref{fig:stability_maps}, \ref{fig:MEGNO}, and \ref{fig:stability_mapsep0}. This is over two orders of magnitude smaller than the shortest relevant timescale, enabling precise resolution of rapid precession and satellite orbital motion without incurring excessive computational cost.
}

\rev{Both the satellite's orbital period and that of the binary star are governed by Kepler's third law. For the binary system, this may be expressed as: 
\[
P_{\mathrm{2}} = 2\pi \sqrt{\frac{a_{2}^{3}}{G (M_{1} + M_{2})}},
\]
where \(a_{2}\) is the semi-major axis of the secondary’s orbit, \(M_{1}\) is the mass of the primary, \(M_{2}\) is the mass of the secondary, and \(G\) is the gravitational constant. The range of orbital periods presented in Table~\ref{tab:dynamical_effects} reflects distances spanning from the Roche limit to 1.5 planetary radii.}

\rev{The Zeipel-Kozai-Lidov (ZLK) effect in the binary system causes oscillations in the planet's inclination and eccentricity due to perturbations from the distant stellar companion. The characteristic timescale for this effect is:
\begin{equation}
P_{\mathrm{ZLK, binary}} = \frac{P_{\mathrm{bin}}^2}{P_{\mathrm{p}}} \frac{M_\star}{M_{\mathrm{bin}}},
\end{equation}
where \(P_{\mathrm{bin}}\) is the orbital period of the binary system, \(P_{\mathrm{p}}\) is the orbital period of the planet, \(M_\star\) is the host star's mass, and \(M_{\mathrm{bin}}\) is the mass of the stellar companion.}

\rev{For the satellite, the Kozai–Lidov effect induces oscillations in its inclination and eccentricity due to the gravitational influence of the host star. The timescale for this process is given by:
\begin{equation}
P_{\mathrm{ZLK, satellite}} = \frac{P_{\mathrm{p}}^2}{P_{\mathrm{s}}} \frac{\Mp}{M_\star},
\end{equation}
where \(P_{\mathrm{s}}\) is the satellite's orbital period.}

\revv{The oblateness of the planet, represented by \(J_\mathrm{2}\), leads to precession of the satellite's orbital plane. The timescale for this effect is:
\begin{equation}
    P_{J_\mathrm{2}} = \frac{2\pi}{\frac{3}{2} J_\mathrm{2} \left( \frac{R_p}{a_s} \right)^2 \sqrt{\frac{G M_p}{a_s^3}}}, 
\end{equation}}
\rev{where \(R_p\) is the planet's radius and \(a_s\) is the semi-major axis of the satellite's orbit. Here we assumed that the satellite's orbital eccentricity \(e_s = 0\) and inclination \(i_s = 0\).}

\rev{General relativistic effects cause the precession of the planet's periapsis. The associated timescale is:
\begin{equation}
P_{\mathrm{rel}} = \frac{2\pi \ap^{5/2}}{3 G^{1/2} \Mp^{3/2} c^2 (1 - e_{\mathrm{p}}^2)},
\end{equation}
where \(\ap\) and \(e_{\mathrm{p}}\) are the planet's semi-major axis and eccentricity, and \(c\) is the speed of light. In addition, relativistic effects can influence the satellite indirectly, through the planet’s periapsis precession, as well as directly via the gravitational gradient associated with its own orbital motion. These influences have seldom been explored, largely because the existence of satellites in such close proximity to their host stars was not anticipated. Employing secular theory to investigate these phenomena could yield valuable insights into their cumulative impact over long timescales, particularly given the broad diversity of temporal scales involved.}

\rev{The ranges of timescales presented in Table~\ref{tab:dynamical_effects} reflect the satellite's orbital distance spanning from the Roche limit to 1.5 planetary radii. These variations illustrate the sensitivity of the dynamical timescales to the satellite's orbital configuration and the underlying physical processes.}
\section{Stability maps}
\label{sec:megno}
\begin{figure}
    \centering
      \includegraphics[width=0.43\textwidth]{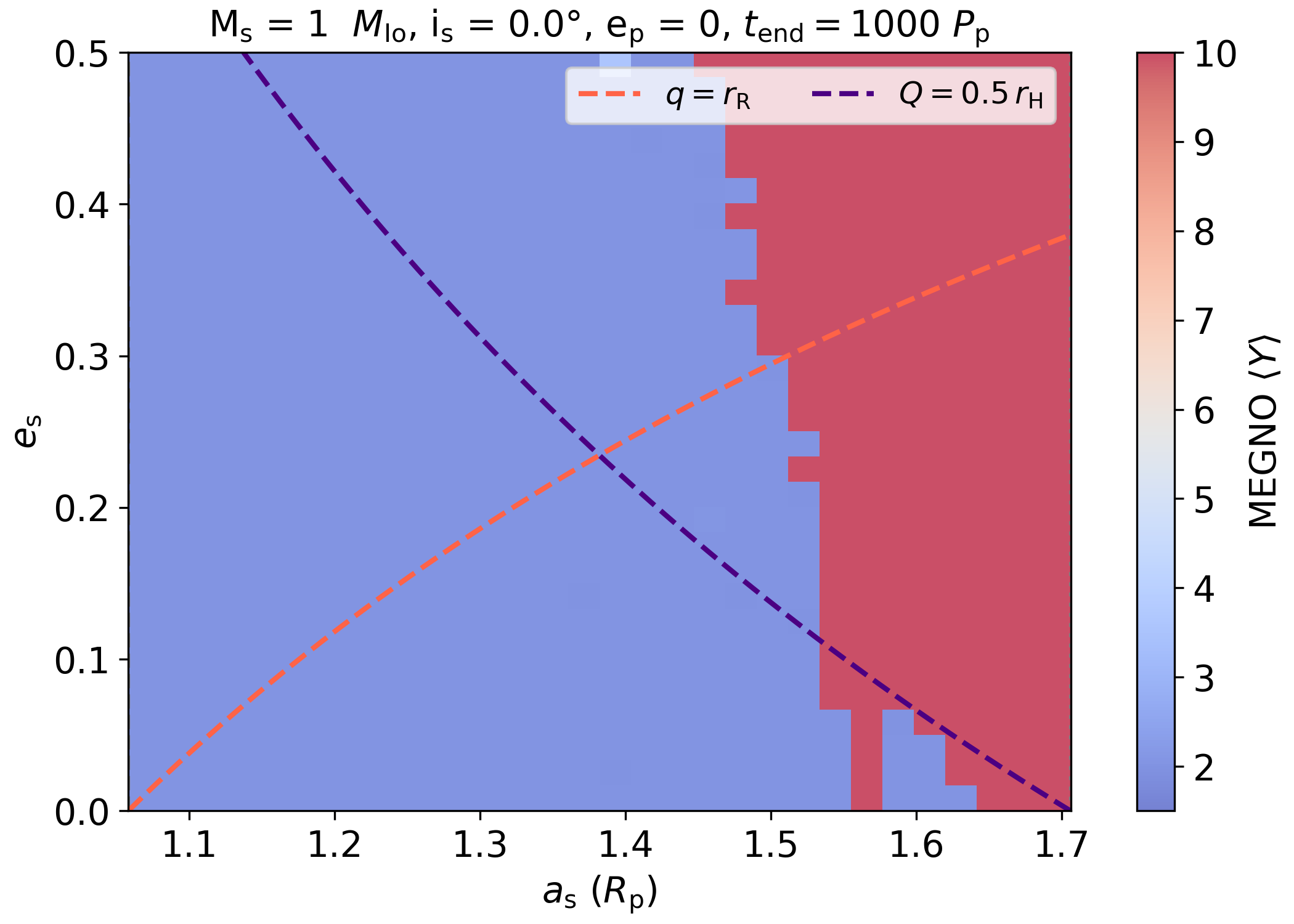}
    \includegraphics[width=0.43\textwidth]{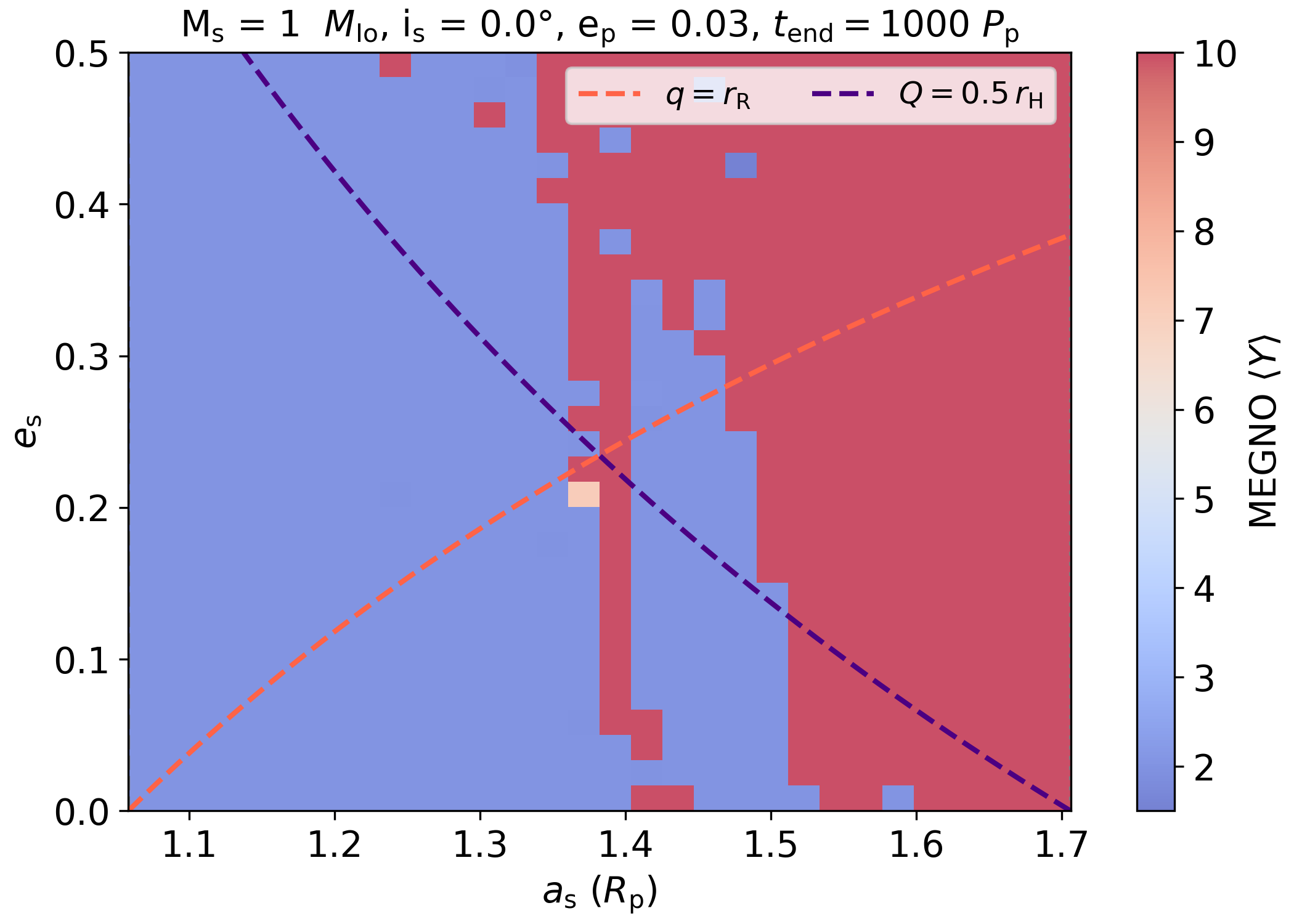}
      \caption{
        \rev{MEGNO stability maps showing regular orbits over 1000 $\Pp$, with key limits marked by the Roche limit (in red) and Hill radius (in black). The top panel assumes a planetary eccentricity of \(e_\mathrm{p} = 0\), while the bottom panel assumes \(e_\mathrm{p} = 0.03\).
        }}

    \label{fig:MEGNO}
\end{figure}

\begin{figure*}
    \centering
    \includegraphics[width=0.44\textwidth]{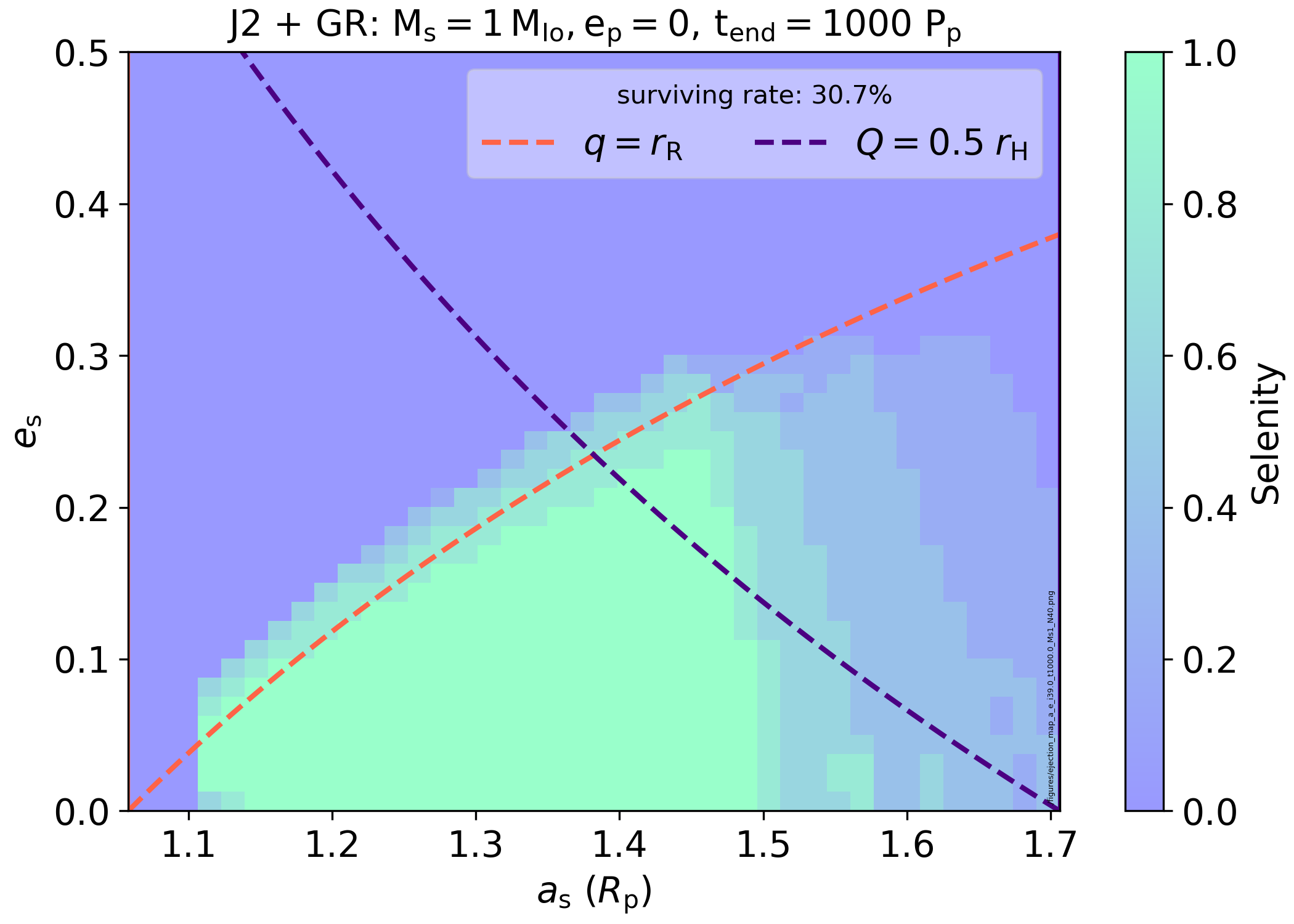}
    \includegraphics[width=0.44\textwidth]{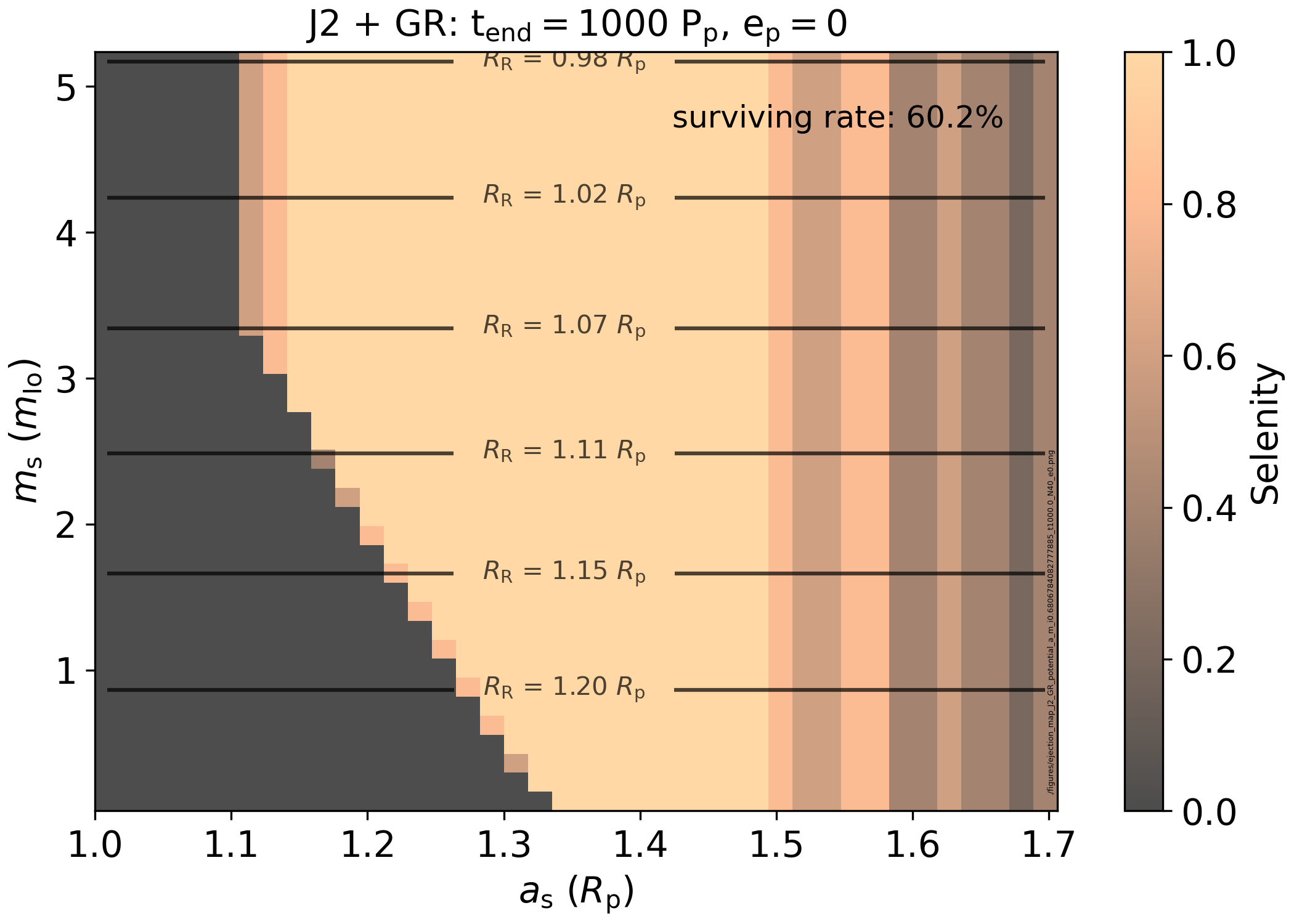}
    \caption{
    \rev{Stability maps showing exomoon survival and ejection probabilities around a close-in giant planet, considering \(J_\mathrm{2}\) and relativistic effects over 1000 $\Pp$ \rev{for a planet with a circular orbit ($e_\mathrm{p}=0$)}. The colour scale represents `Selenity’, indicating the likelihood of moons remaining bound. Left: Stability in the \(a_\mathrm{s} - e_\mathrm{s}\) plane, with the Roche limit and Hill radius marked. Right: Stability in the \(a_\mathrm{s} - M_\mathrm{s}\) plane, highlighting variations with satellite mass and planet distance.
    }}
    \label{fig:stability_mapsep0}
\end{figure*}

\rev{The MEGNO stability maps shown in Fig.~\ref{fig:MEGNO} illustrate the dynamical behaviour of hypothetical moons around WASP-49Ab, focusing on the coplanar configuration. When the satellite’s eccentricity is set to zero (upper panel in Fig.~\ref{fig:MEGNO}), no instability gap is observed, emphasising the stabilising role of a circular orbit. However, when the planet’s eccentricity is non-zero (lower panel in Fig.~\ref{fig:MEGNO}), a distinct instability gap appears near \(1.4 \, \Rp\), which shifts to larger semi-major axes as the satellite’s inclination increases in other maps (not shown). This gap is not associated with mean-motion resonances but rather arises from the planet’s eccentricity, which induces secular perturbations that destabilise the satellite’s orbit in specific regions.}

\rev{The simulations were conducted using a timestep of \(10^{-5} \, \mathrm{yr}\), ensuring precise resolution of the satellite's orbital dynamics. Each simulation covered 1000 planetary orbits, equivalent to nearly \(10^5\) satellite periods, providing a comprehensive assessment of the system's stability over long timescales.}

\revv{The MEGNO indicator used here relies on a Newtonian point-mass model, which excludes effects such as planetary \(J_\mathrm{n}\) zonal deformation and relativistic corrections. These limitations stem from its implementation in codes like {\tt REBOUND}, rather than the MEGNO framework itself. In principle, MEGNO can extend beyond classical N-body dynamics, provided the equations of motion incorporate conservative forces. Examples include the Schwarzschild relativistic correction, the conservative tidal distortion potential, and the \(J_\mathrm{2}\) zonal deformation effect.}

\revv{Moreover, MEGNO is compatible with non-autonomous systems, where time explicitly appears as a variable in the equations of motion. This has been demonstrated in studies of systems with effective 1 degree of freedom, derived from three coordinates and two constraints \footnote{Such as those explored by Carles Simó, see e.g.,  \url{https://bgsmath.cat/people/?person=carles-simo}}. For the dynamical maps presented, Newtonian forces dominate the short-term dynamics, justifying the first-order approximation. Comparisons with the “selenity” factor further validate this approach as a robust representation of the core dynamics, despite minor unmodelled perturbations.
However, as shown in the MEGNO and selenity maps in Fig.~\ref{fig:stability_maps}, Fig.~\ref{fig:MEGNO} and Fig.~\ref{fig:stability_mapsep0}, the instability near \(1.4 \, \Rp\) persists even when these additional effects are included, indicating that it is a robust dynamical feature of the system. This instability does not appear when the planet’s eccentricity is set to zero (\(e_\mathrm{p}=0\)), suggesting that it is primarily driven by gravitational and secular perturbations related to the planet's non-zero eccentricity (\(e_\mathrm{p} \neq 0\)). These perturbations cause oscillations in the satellite’s orbital elements, particularly eccentricity, potentially leading to destabilizing scenarios like perigee near the Roche limit or apogee beyond the secondary Hill radius. The recurrence of this instability in both MEGNO and selenity maps suggests a secular resonance, possibly \(\nu\)-type, though confirmation requires detailed secular theory analysis.
}

\rev{ To further investigate the role of satellite mass on stability, we conducted additional simulations for \(\Ms = 0.1 \, M_\mathrm{Io}\) and \(\Ms = 0.01 \, M_\mathrm{Io}\). The survival rates decreased markedly, from 25\% for \(\Ms = M_\mathrm{Io}\) to 17\% and 16\%, respectively. This finding confirms that satellite mass plays a significant role in the system’s dynamics, as smaller masses reduce the satellite's gravitational influence and its ability to resist destabilising perturbations. These effects are particularly apparent near \(1.4 \, R_p\), where the interplay between the planet’s eccentricity-driven secular perturbations and the satellite's reduced stabilising influence leads to enhanced instability. }

\rev{This result strengthens the trends observed in Fig.~\ref{fig:stability_maps}, indicating that satellites with masses below \(M_\mathrm{Io}\) are increasingly prone to destabilisation. Such findings suggest a practical lower limit on the masses of stable satellites in systems like WASP-49Ab, highlighting the crucial role of satellite mass in long-term dynamical stability.
}

\end{document}